*sensors*

*Article*

# Reverse Engineering and Security Evaluation of Commercial Tags for RFID-Based IoT Applications

**Tiago M. Fernández-Caramés \*, Paula Fraga-Lamas, Manuel Suárez-Albela and Luis Castedo**

Department of Electronics and Systems, Faculty of Computer Science, Universidade da Coruña,
15071 A Coruña, Spain; paula.fraga@udc.es (P.F.-L.); m.albela@udc.es (M.S.-A.); luis.castedo@udc.es (L.C.)
\* Correspondence: tiago.fernandez@udc.es; Tel.: +34-981-16-7000 (ext. 6088)

**Abstract:** The Internet of Things (IoT) is a distributed system of physical objects that requires the seamless integration of hardware (e.g., sensors, actuators, electronics) and network communications in order to collect and exchange data. IoT smart objects need to be somehow identified to determine the origin of the data and to automatically detect the elements around us. One of the best positioned technologies to perform identification is RFID (Radio Frequency Identification), which in the last years has gained a lot of popularity in applications like access control, payment cards or logistics. Despite its popularity, RFID security has not been properly handled in numerous applications. To foster security in such applications, this article includes three main contributions. First, in order to establish the basics, a detailed review of the most common flaws found in RFID-based IoT systems is provided, including the latest attacks described in the literature. Second, a novel methodology that eases the detection and mitigation of such flaws is presented. Third, the latest RFID security tools are analyzed and the methodology proposed is applied through one of them (Proxmark 3) to validate it. Thus, the methodology is tested in different scenarios where tags are commonly used for identification. In such systems it was possible to clone transponders, extract information, and even emulate both tags and readers. Therefore, it is shown that the methodology proposed is useful for auditing security and reverse engineering RFID communications in IoT applications. It must be noted that, although this paper is aimed at fostering RFID communications security in IoT applications, the methodology can be applied to any RFID communications protocol.

## 1. Introduction

The Internet of Things (IoT) is paving the way for revolutionizing the way we interact with daily objects. Such a paradigm presents itself as a network of smart objects that collect information through sensors, execute physical actions through actuators, and exchange data with other devices through communication interfaces. In order to carry out such tasks, the smart objects have to be able to identify themselves and be identified by others.

Among the different technologies to perform identification, RFID (Radio Frequency Identification) is currently one of the best positioned, since it has been proven successful for identifying elements in multiple practical applications, like animal identification [1], healthcare [2], passport control [3], transportation [4], supply chain traceability [5,6], maritime freight container tracking [7], protective equipment verification [8], or toll payments [9,10]. There are already a number of RFID-based developments explicitly designed for IoT applications in different areas, like logistics [11], ITS (Intelligent Transportation Systems) [12], smart environments [13], parking systems [14], defense and public safety [15], assisted-living [16], smart power sockets [17] or vehicle tracking [18].

These are just some examples, since the list of possible RFID-based IoT applications is huge (Figure 1 shows some of the most popular).

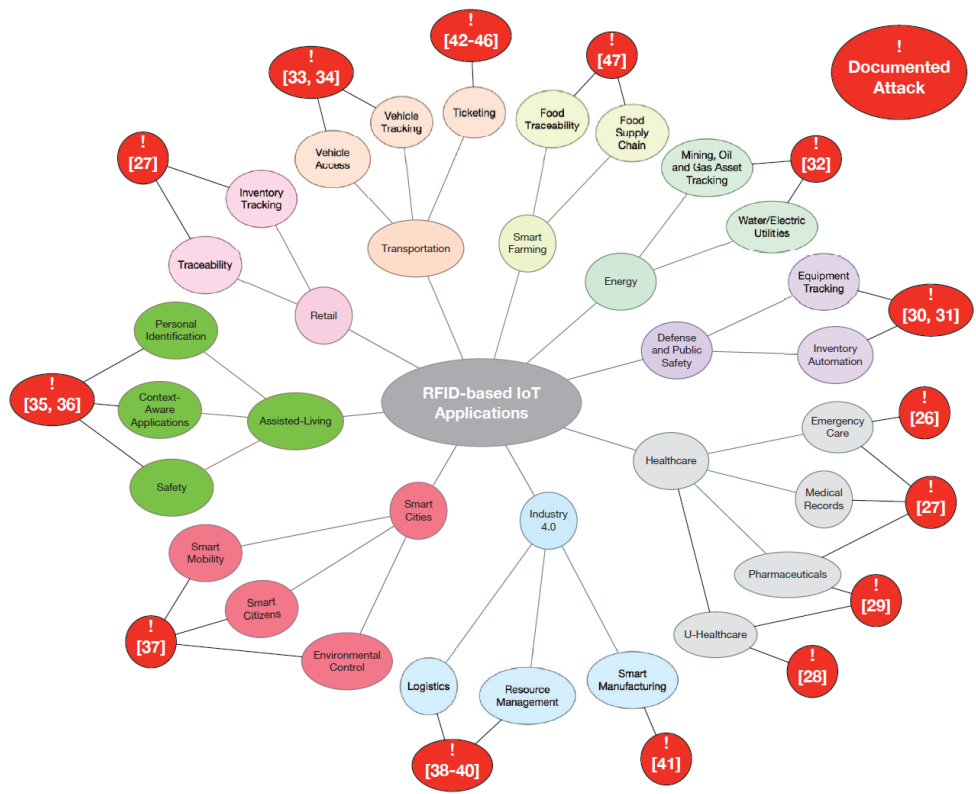

**Figure 1.** Main RFID-based IoT applications.

Nevertheless, despite RFID's popularity, many developers of applications have neglected security: it is easy to find commercial systems that contain critical security flaws and vulnerabilities that allow for cloning tags or for straight signal replaying [19,20]. Such vulnerabilities let attackers access certain services or facilities, get or alter personal information, and even track users.

Additionally, many RFID systems are susceptible to reverse engineering. Thus, certain hardware and software components can be extracted and analyzed in order to reproduce them. For instance, recently, multiple authors have been able to emulate communications protocols and reverse-engineer cryptographic algorithms using, in most cases, low-cost equipment [21–25]. Figure 1 contains references to the latest analysis about possible attacks on RFID in IoT applications for specific fields like healthcare [26–29], defense and public safety [30,31], energy [32], vehicle access [33,34], retail [27], assisted-living [35,36], smart cities [37], Industry 4.0 [38–41], ticketing [42–46], or smart farming [47].

Countermeasures can be taken to prevent attacks. The most common defenses include the use of cryptography [48], automatic malware detection [49], improving resistance to cloning [50], uncovering rogue devices [51], or the use of secure authentication schemes [52]. However, it is common to find commercial RFID systems that due to cost or speed have such security features disabled. Unfortunately, it is even more common to identify already-broken RFID security systems still in use. To tackle such issues this paper proposes a methodology, whose initial version was described in [53], that allows developers to detect the most common RFID security flaws. Note that this paper is aimed at fostering security in RFID-based IoT applications, but the methodology proposed can be applied to any RFID system.

Thus, the contribution of this article is threefold. First, in order to establish the basics, it gives a detailed and easy to follow review of the most common flaws found in RFID-based IoT systems, including the latest attacks described in the literature. Second, a step-by-step methodology that eases the detection and mitigation of RFID security vulnerabilities is presented (we are not aware of any similar methodology in the literature). Third, the most recent RFID security tools are analyzed and the methodology proposed is applied through one of them (Proxmark 3) in real-world scenarios.

The rest of this article is organized as follows. Section 2 enumerates the main types of attacks on RFID and the latest hardware tools to test RFID security. Section 3 describes the methodology proposed for auditing RFID security and reverse engineering communications protocols. In Section 4 the methodology is applied to three different tags: an access control card, a university card, and an animal identification tag. Finally, Section 5 is devoted to conclusions.

## 2. Related Work

### 2.1. Basic Types of RFID Systems

Before describing the tools necessary to analyze communications protocols and RFID security, we enumerate briefly the main types of RFID systems existing on the market. The following information is well-known, but we consider that is necessary to mention it briefly for the sake of clarifying the basic concepts related to the methodology and the hardware available. A detailed description of the principles that regulate how RFID works is out of the scope of this article, but the interested reader can get a good overview of the technology and its basic security implications in [54].

Depending on the frequency band, the following RFID systems can be found:

- LF (Low Frequency) RFID. According to the ITU (International Telecommunications Union), the LF band goes between 30 kHz and 300 kHz. Frequency and power in this band are not regulated globally in the same way: most systems operate at 125 kHz, but there are some at 134 kHz. The reading range provided is short (generally up to 10 cm), so, in practice, LF devices are not usually sensitive to radio interference. Its most popular applications are access control and animal identification (mainly for pets and livestock).
- HF (High Frequency) RFID. Although the HF band goes from 3 MHz to 30 MHz, most systems operate at 13.56 MHz. HF systems can reach a reading distance of up to 1 m, what can lead to interference and, therefore, MAC (Medium-Access Control) mechanisms have to be implemented. This sort of RFID systems is massively used in transportation, payment, ticketing, and access control.
- UHF (Ultra-High Frequency) RFID. The UHF band actually covers from 300 MHz to 3 GHz, but most systems operate in the ISM (Industrial-Scientific-Medical) bands around 860–960 MHz and 2.45 GHz. UHF tag can be easily read at 10 m, so they are ideal for inventory management and item tracking in logistics.

All these RFID systems can also be classified according to the way the tags are powered:

- Passive systems. They do not need internal batteries to operate, since they rectify the energy sent through the reader's antenna. There are LF, HF, and UHF passive systems, which nowadays can be easily read at a 10-m distance.
- Active systems. They include batteries, what allow them to reach further distances (usually up to 100 m). Due to power regulations, almost all active systems operate in the UHF band.
- Semi-active, semi-passive or BAP (Battery-Assisted Passive) systems. They decrease power consumption by using batteries just for powering the tags for certain functionality. Commonly, batteries are used to power up the basic electronics, while the energy obtained from the reader is used for powering the communications interface.

*2.2. Attacks against RFID*

2.2.1. Risks and Threats

Information security threats have been traditionally classified according to what is known as the CIA Triad:

• Confidentiality. It is related to the importance of protecting the most sensitive information from unauthorized access.
• Integrity. It consists in protecting data from modification or deletion by unauthorized parties, and ensuring that, when authorized people make changes, they can be undone if some damage occurs.
• Availability. It is the possibility of accessing the system data when needed.

If any of these three principles is not met, then security can be said that it has been broken.

Like other technologies, RFID is exposed to security threats and, specifically, to attacks on the confidentiality, integrity and availability of the data stored on the tags, or on the information exchanged between a reader and a tag. When these threats are associated with the probability of occurrence of an event that causes damage to an informational asset, they are known as risks. Two kinds of risks can be basically distinguished:

• Security risks. They are derived from actions able to damage, block or take advantage from a service in a malicious way. The action is usually carried out with the objective of obtaining a profit or just to damage the access to certain service.
• Privacy risks. These risks affect the confidential information of the users. In some cases, when a user interacts constantly with the environment, objects, and people around, an attacker would be even able to obtain extremely accurate information on the personal data, location, behavior and habits. In the case of RFID, there are mainly two privacy risks:

    – Unauthorized access to personal data. Many systems store private data on RFID tags or transmit them when a tag and a reader exchange information.
    – Personal tracking. This is probably the most feared, since an attacker might determine routes, purchases and habits of a specific person.

In real life, most risks are a mixture of both security and privacy risks: they threaten RFID security in order to get access to the information stored or to the data exchanged in a transaction.

2.2.2. Physical Attacks

This type of threat consists in using some kind of physical medium to attack a tag or the RFID communications. There are mainly five basic attacks:

• Reverse engineering. Most tags are not tamper-proof and can be disassembled and analyzed. A description of the most relevant reverse-engineering attacks is given later in Section 2.4.
• Signal blocking or jamming. It consists of blocking tag communications to avoid sending data to a reader.
• Tag removal. It consists of removing an RFID tag or replacing it with another one.
• Physical destruction. In this case the attacker destroys the RFID tag by applying pressure, tension loads, or high/low temperatures; by exposing the tag to certain chemicals; or by just clipping the antenna off.
• Wireless zapping. RFID zappers are able to send energy remotely that, once rectified, is so high that certain components of the tag might be burned.

2.2.3. Software Attacks

These attacks are related to software bugs or vulnerabilities found in tags or in the RFID reader. The most common are:

- Remote switch off. Researchers have found that it is possible to misuse the kill password in some tags (EPC Class-1 Gen-2) with a passive eavesdropper and then disable the tags [55].
- Tag cloning. In this attack, the Unique Identifier (UID) and/or the content of the RFID is extracted and inserted into another tag [56].
- Command injection. Some readers are vulnerable to remote code execution by just reading the content of a tag [57].
- SQL injection. It has been found that some reader middleware is vulnerable to the injection of random SQL commands [57].
- Virus/Malware injection. Although difficult to perform in the vast majority of RFID tags due to their low storage capacity, it is possible in certain tags to insert malicious code that is able to be transmitted to other tags [57].
- Network protocol attacks. Many systems integrate back-end databases and connect to networking devices, which are susceptible to the same vulnerabilities as any other general purpose networking device.

2.2.4. Channel Attacks

Channel attacks refer to threats related to the lack of security on the communications between the reader and the tag. The following are the most popular attacks:

- Unauthorized reading. Most RFID tags can be easily read without leaving a trace, although readings are limited to relatively short distances. Some of the latest measures to prevent this kind of attacks make use of sophisticated techniques [58,59].
- Denial of Service (DoS) attacks. The channel is flooded with such a large amount of information that the reader cannot deal with the signals sent by real tags [60].
- Signal replaying. It consists in recording the RFID signal in certain time instants with the objective of replaying it later.
- Man-in-the-Middle (MitM) attacks. They consist in placing an active device between a tag and a reader in order to intercept and alter the communications between both elements [58,61].
- Relay/amplification attacks. They consist in amplifying the RFID signal using a relay, so the range of the RFID tag is extended beyond its intended use [33,62].

*2.3. Countermeasures Against the Most Common Attacks*

RFID systems can take one or more the following measures against the attacks previously described:

- Reader-Tag authentication. Both devices should carry out a two-way authentication, so only legitimate devices can communicate. This mechanism prevents certain types of remote tag destruction (i.e., only an authorized user can send a kill command), unauthorized readings, and MitM attacks. For instance, a lightweight authentication protocol is presented in [63].
- Rogue device detection. If a reader is provided with the capacity of detecting abnormal tag behaviors, it might avoid DoS attacks, certain unauthorized readings, command/virus/malware injection, and network protocol attacks.
- Use of cryptography. Due to the limited power and performance of most RFID tags, complex cryptography is not usual in most reader-tag communications. However, a minimum level of communications confidentiality has to be provided by RFID systems, so the most relevant information should be encrypted. Basic cryptography can prevent eavesdropping, MitM attacks, and unauthorized readings.

- Data integrity verification. RFID systems should ensure that the data received has not been tampered or modified by an attacker. This verification is key in MitM attacks.

*2.4. Reverse-Engineering Attacks*

In this paper, a methodology that may be used for auditing security and reverse-engineer communications protocols in RFID systems is described. In the latter case is important to emphasize that there are different alternative attacks that researchers have tested in the last years:

- Communications protocol analysis. This is related to channel attacks: the communications between the reader and the tags are captured and analyzed. It is probably the most popular attack because it is non-intrusive and the cost of the hardware is relatively low in comparison to other attacks. For instance, an example of a communications protocol analysis is described in [25]: the authors detail how they reverse engineered and emulated an LF tag for sport events with the help of an Arduino board and a few electronic components. However, note that it is quite difficult to derive all of the functionality, specially in the case of encrypted and obfuscated communications. Although cryptography hinders communications protocols analysis, it is not actually implemented in many tags, since additional hardware (i.e., higher economic cost) and power are required, and communications latency is increased. The methodology proposed in this paper is actually aimed at performing communications protocol analysis.
- Power analysis. It is a type of non-intrusive attack that assumes that power consumption (or the electromagnetic field) is related to the execution of certain instructions. A good description of how to carry out a power analysis is presented in [23,24], where the authors attack different commercial HF and UHF RFID tags. A remarkable work is also [21], that describes what the author claims to be the first remote power analysis against a passive RFID tag. To prevent power analysis, the different functions to protect must be designed to consume the same amount of power: although the algorithms may seem inefficient, the attacker would not distinguish between the different processes.
- Optical analysis. This attack is widely used for reverse-engineering microchips and, therefore, it can be used for studying the internal hardware of an RFID tag. Before performing such an analysis, the external enclosure has to be removed, which involves using acid and, less frequently, a laser beam. Then, an optical or electron microscope can be used to analyze the hardware. An excellent example of optical analysis is described in [22], where the authors detail how they reverse-engineered the security of MIFARE Classic cards (the authors first performed an optical analysis, and then a communications protocol analysis). To avoid reverse engineering through optical analysis, the designers of RFID tags can embed non-functional logic to misguide the attackers, re-position the internal hardware to make the analysis more difficult, or implement certain key functionality in software instead of hardware.
- Electronic analysis. It is usually performed in combination with optical analysis to get a better picture on how an RFID circuit works. It consists in applying really small probes to read or induce voltages in different parts of the chip when carrying out certain operations. Bus obfuscation and communications encryption are usually effective against this kind of analysis.

*2.5. Hardware Tools for Auditing RFID Security and Reverse Engineering Communications Protocols*

In recent years, a number of projects have been developed with the aim of facilitating researchers low-level access to RFID communications. Some of them are just software tools that can be used with commercial RFID readers [64], while others involve specific hardware [39,65–69] or certain firmware [70]. Hardware developments are specially interesting: some devices can emulate readers [66,67], others can emulate just tags [39,68], and a few can emulate both kinds of devices [65,69].

RFIDIOt [64] is a set of open-source software tools developed as python libraries aimed at analyzing RFID devices. These libraries are compatible with different High Frequency (HF) and Low Frequency (LF) readers (manufactured by ACG, Omnikey or Frosch Electronics), and support reading/writing to multiple tags (e.g., MIFARE, SLE, ISO/IEC 14443-A, ISO/IEC 14443-B, ISO/IEC 15693, ISO 18000-3, NFC, ICODE, EM 4x tags, Hitag, or TI-RFID).

Tastic [66] focuses on reading LF and HF tags at a long distance (up to one meter). It specifically targets badge systems like HID Prox, Indala Prox or HID ICLASS. It is based on an Arduino board that connects to standard DATA0/DATA1 Wiegand outputs.

OpenPCD [67] is an open-source and open-hardware system able to emulate and sniff data from HF RFID/NFC cards (e.g., ISO/IEC 14443, ISO/IEC 15690, MIFARE, ICLASS). It supports the libNFC library [71] and has been designed around NXP's PN532, which is a transmission module that embeds a 80C51 microcontroller with 40 KB of ROM and 1 KB of RAM.

OpenPICC [68] is the counterpart of OpenPCD: it emulates HF tags like the ones compliant with ISO/IEC 14443 and ISO/IEC 15690. It is based on a 32-bit ARM microcontroller (AT91SAM7S256) with 128 KB of flash memory and 64 KB of SRAM.

There are not many academic platforms developed to test RFID security. One good example is described in [39]. Such a platform is composed by a microcontroller and an Field-Programmable Gate Array (FPGA). Its aim is to evaluate HF and Ultra-High Frequency (UHF) RFID tags. The latest academic development as of writing is the Chameleon Mini [69], which has been promoted by the Ruhr University (Bochum, Germany): it is a versatile RFID tag emulator compliant with ISO/IEC 14443 and ISO/IEC 15693 (for instance, it currently supports MIFARE Classic 1K/4K/Ultralight emulation).

The platform selected in this paper to analyze RFID security is Proxmark 3 [65], which is an open-source system able to transmit at LF (125-134 kHz) and HF (13.56 MHz). The system contains an Atmel AT91SAM7S256 (256 KB of Flash and 64 KB of RAM), an FPGA (Xilinx Spartan-II) and an 8-bit Analog-to-Digital Converter (ADC). It is powered through an USB and has an SV2 connector for the antenna, which contains four pins: two are for the HF antenna, and the other two are for the LF antenna. All these components can be observed in Figure 2. The Proxmark 3 was our choice to test commercial RFID systems because of its main features:

- It operates in HF and LF, where most popular RFID applications work (e.g., identification tags, payment cards or passports). This is due to the hardware cost in such frequency bands and because the reading distance is enough for the applications. UHF is also heavily used in other fields, where more reading distance is required (e.g., logistics), and tags have become inexpensive, but the reading hardware (i.e., readers, antennas, muxes and amplifiers) is more expensive than most LF and HF devices.
- Its ability to sniff easily communications between a reader and different tags.
- The possibility of emulating diverse RFID communications protocols. The official firmware supports some basic protocols, but it is relatively easy to develop and upload new code to the embedded ARM microcontroller and to its FPGA.
- The community behind Proxmark 3, which has been extending the official firmware to add new features.

When the Proxmark 3 acts as an RFID receiver, the signal that comes from the antenna goes through the ADC and is converted from analog to digital. Then, the digital data are sent through an 8-bit bus to the FPGA, where they are demodulated. Finally, the signal is sent from the FPGA to the microcontroller through the SPI to deal with the RFID protocol. When the Proxmark acts as a transmitter, the same steps are performed but in reverse order. The FPGA modulators/demodulators are developed in Verilog, while the Atmel microcontroller is programmed in C. There is also a client application developed in C able to send remote commands to interact with the device. Different custom firmwares have been developed for Proxmark 3. An example is Proxbrute [70], created by McAffee in order to extend Proxmark functionality to perform brute force attacks, mainly against access control systems.

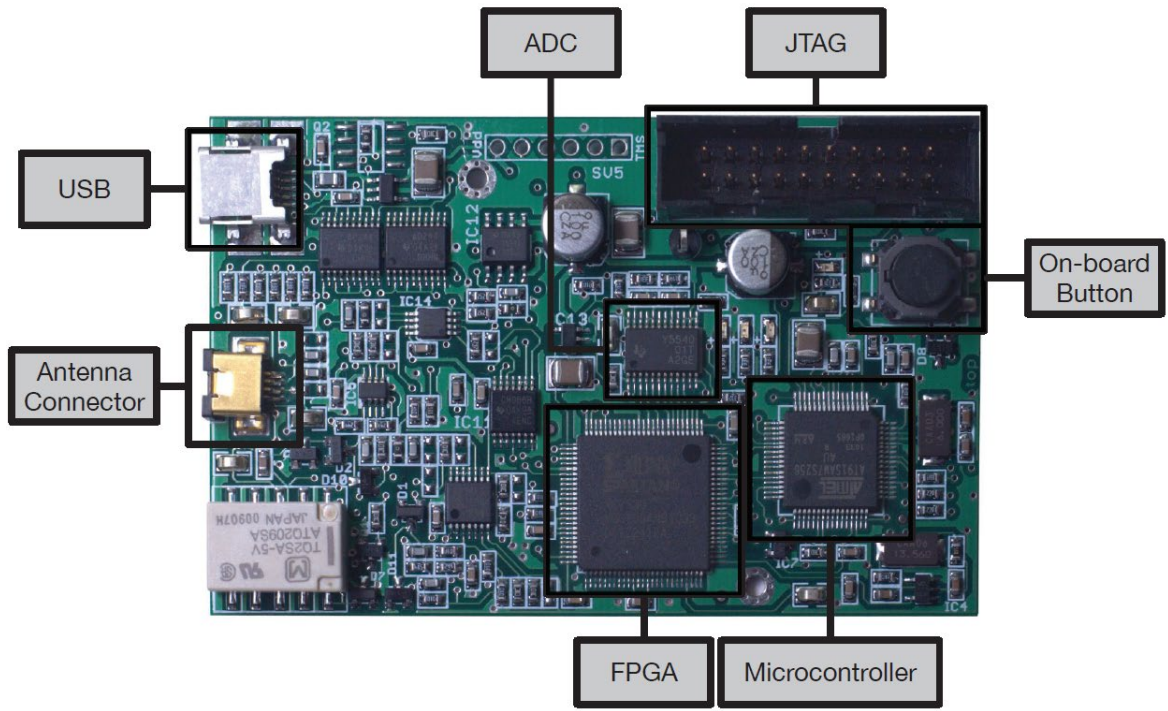

**Figure 2.** Main components of Proxmark 3.

## 3. A Methodology for Reverse Engineering Communications Protocols and Auditing RFID Security

### 3.1. Objectives of the Methodology

The methodology presented in the next subsection was initially devised for testing the security claims of one manufacturer of a commercial system used daily for transportation by more than 200,000 users [53]. After verifying that the non-documented communications protocol of such an application had multiple vulnerabilities, we continue to check other commercial RFID systems.

It must be emphasized that RFID penetration tests are valid for evaluating the security of a product, but reverse engineering approaches are able to expose the internal structure and reveal vulnerabilities. These vulnerabilities include the existence of high-privilege modes, debugging functionality or backdoors implemented intentionally by the manufacturer.

Note that the knowledge obtained by applying the methodology proposed might also be used for other purposes: the gathering of information on poorly or non-documented RFID systems, the replication of software copyrighted without violating the law, or for espionage purposes. In this latter case a company may reverse-engineer a competing product to study its inner workings and estimate the hardware cost in order to enhance its own products and determine if it is possible to offer better prices.

It must be indicated that we are only aware of one other methodology focused on reverse engineering RFID systems [72]: the one used by RIDAC [73], an open-source framework for auditing RFID security released by Oulu university (Finland) in 2009. The methodology has similar objectives, but it is structured in processes instead of steps, and is oriented towards the specific use of RIDAC software.

### 3.2. Basic Steps

With the objective of automating the reverse engineering process and the security audit of commercial RFID systems, a methodology was devised. Such a methodology first determines the most relevant parameters of a tag (i.e., operating frequency, coding scheme, and modulation), and then identifies its RFID standard (or tries to reverse engineer the communications protocol and the internal data structure).

The methodology flow diagram is depicted in Figure 3, where the following main steps can be observed:

- Visual inspection. Before analyzing the characteristics of a tag, it is first recommended to look for external signs that might indicate the manufacturer, the model, or the RFID standard. If any of such data is recognized, it is usually straightforward to obtain the basic parameters and details on the communications protocol.

- FCC (Federal Communications Commission) ID or equivalent. One of the most relevant external signs for determining the internal parameters of a tag is its FCC ID (or its equivalent in other parts of the world). The FCC is an agency of the United States that regulates radio communications. Each FCC-approved radio device receives a unique FCC ID that must be marked permanently and has to be visible to the buyer at the time of purchase. Such an FCC ID is composed by 4–17 alphanumeric characters. The first three characters are the Grantee Code, which identifies the company that asks for the authorization of the radio equipment. The rest of the characters (between 1 and 14) are the Product Code. If there is an FCC ID label on an RFID tag or on a reader, it is possible to obtain through the FCC ID search page [74] information like the name of the company that has applied for the authorization, the lower and upper operating frequencies, block diagrams, schematics, and even external/internal photos of the device.

- Frequency band detection. In most commercial systems it is not common to show external clues about the characteristics of a tag, so, in these cases, a detailed analysis has to be carried out. The first parameter to determine is the operation frequency of the tag. Most tags use LF, HF or UHF bands. If through the previous steps of the methodology it is not possible to determine the frequency, two additional processes can be performed:

    - Disassemble and analyze the hardware. This step aim is to study the internal components in order to determine the operation frequency. The most interesting elements are the ones related to the radio interface: transceivers, amplifiers, crystals, and filters allow us to determine the operation band and then estimate the frequency. In this case, transceiver datasheets are the fastest way to obtain an accurate frequency value. RFID readers are usually really easy to disassemble, but RFID tags require more sophisticated tools and techniques. An excellent description on how to disassemble and analyze an RFID tag is given in [22]. In such a paper the authors first use acetone to dissolve the external plastic encapsulation and isolate the blank chips. This process requires about half an hour, and it is easier and safer than other alternatives like the use of fuming nitric acid. Next, each layer of the silicon chips is removed through mechanical polishing (e.g., by using sandpaper), since it is easier to control than chemical etching. Note that very fine grading is required (e.g., 0.04 μm), because the layers can have a thickness of around a micrometer. Finally, after a successful polishing, the internal circuitry can be usually analyzed through a standard optical microscope.

    - Radio spectrum analysis. In this case, a spectrum or network analyzer, or an oscilloscope, is used to detect the operation frequency. The objective of such an analysis is to determine the resonant frequency of a passive RFID tag. The process models the RFID tag as a simple RLC parallel resonant circuit, what allows for obtaining the resonant frequency easily through Thomson equation:

$$f_r = \frac{1}{2\pi\sqrt{LC}}$$

where $L$ is the inductance and $C$ is the capacitance. The key for measuring the resonant frequency is the fact that the impedance of the measuring antenna, the reflection coefficient (that measures how much of an electromagnetic wave is reflected by an impedance discontinuity), and the transmission coefficient (that measures how much of an electromagnetic wave passes through a surface) change significantly at frequencies in the vicinity of the RFID tag resonant frequency, $f_r$. Therefore, the resonant frequency can be determined by scanning a frequency range and observing when these changes reach their peak. The whole process varies depending on the measurement equipment used, but some manufacturers ease it by offering step-by-step tutorials [75].

There is also a cheaper option for carrying out this analysis that involves working with SDR (Software-Defined Radio) tools like the ones cited in Section 2.5, which can be reprogrammed to be used as spectrum analyzers. For instance, the USRP platform [76] has been proposed recently for spectrum monitoring [77] and sensing in cognitive radio applications [78,79], what can be re-purposed for RFID transmission frequency detection.

- LF/HF tag parameter analysis. If it is verified that the RFID system is LF or HF, the next step of the methodology requires determining the modulation and the coding scheme used by the tag. These tasks involve the use of the appropriate tool to perform a detailed analysis of the radio signals. Such a tool may be a bench oscilloscope with a measuring antenna or similar hardware (e.g., Proxmark 3) that allows for acquiring the RFID signals and then showing the wave received through a display. Thus, the identification is mainly visual, so the analysis becomes easier when the researcher has experience on recognizing the most common modulation and coding patterns. There also exists the possibility of using automatic recognition algorithms, which have been used for a long time (mainly in the military field) [80], but, very recently, they have been updated and improved to detect RFID physical layer characteristics [81,82].

- UHF tag parameter analysis. In the case of RFID UHF systems, the study can become difficult because, although most passive tags are compliant with the EPC Gen 2 standard, there are a number of companies that make use of proprietary protocols. In such a case, reverse engineering may require using SDR platforms like USRP, MyriadRF [83] or HackRF One [84] to study and then emulate the RFID communications protocol. In the case of the USRP platform, several researchers have presented really good references on how to implement USRP-based systems for identifying UHF tags over the last years [85,86].

- Standard analysis. Once the frequency, the modulation, and the coding scheme have been obtained, it is straightforward to determine whether there exists an RFID standard compliant with such a configuration. If there is not, the research may involve reverse engineering a proprietary protocol. However, due to compatibility purposes, most massively commercialized LF, HF and UHF tags follow well-known RFID standards. Table 1 provides a fast way to determine the RFID standard from the frequency, modulation and coding previously determined. Such a Table shows the wide variety of implementations, which include modulations like Amplitude-Shift Keying (ASK), Double-Sideband ASK (DSB-ASK), Single-Sideband ASK (SSB-ASK), Phase-Reversal ASK (PR-ASK), Frequency-Shift Keying (FSK), Binary-Phase Shift Keying (BPSK), Differential BPSK (DBPSK), Phase-Jitter Modulation (PJM), On-Off Keying (OOK) or Gaussian Minimum Shift Keying (GMSK); and coding schemes like Differential Bi-Phase (DBP), Dual Pattern (DP), Non-Return-to-Zero (NRZ), Non-Return-to-Zero-L (NRZ-L), Pulse-Interval Encoding (PIE), Manchester, Pulse-Position Modulation (PPM), Modified Frequency Modulation (MFM), modified Miller, or FM0.

**Table 1.** Physical layer characteristics of the most relevant RFID standards.

| Standard | Mode/Type | Communications | Carrier Frequency | Modulations Supported | Coding Schemes | Main Applications |
|---|---|---|---|---|---|---|
| ISO/IEC 11785 | FDX/FDX-B | - | 134.2 kHz | ASK | DBP | Animal |
| | HDX | - | 134.2 kHz | FSK | NRZ | identification |
| ISO/IEC 14223 | FDX/HDX-ADV | - | 134.2 kHz | ASK | PIE | Advanced animal tagging |
| ISO/IEC 18000-2 | Type A | Reader to Tag | 125 kHz | ASK | PIE | Smart cards, ticketing, |
| | | Tag to Reader | 125 kHz | ASK | Manchester, DP | animal identification, |
| | Type B | Reader to Tag | 125 kHz or | ASK | PIE | factory data collection |
| | | Tag to Reader | 134.2 kHz | FSK | NRZ | |
| ISO 21007 (LF) | - | - | 125 kHz | ASK | Manchester | Identification of gas cylinders |
| ISO/IEC 18000-3 | Mode 1 | Reader to Tag | 13.56 MHz | DBPSK | PPM | Smart cards, |
| | | Tag to Reader | 13.56 MHz | DBPSK | Manchester | small item management, |
| | Mode 2 | Reader to Tag | 13.56 MHz | PJM | MFM | libraries, transportation, |
| | | Tag to Reader | 13.56 MHz | BPSK | MFM | supply chain, passports, anti-theft |
| | Mode 3 | Mandatory Mode | 13.56 MHz | ASK | PIE | |
| | | Optional Mode | 13.56 MHz | PJM | MFM | |
| ISO/IEC 15693 | - | Reader to Tag | 13.56 MHz | ASK | PPM | Vicinity cards and |
| | | Tag to Reader | 13.56 MHz | ASK or FSK | Manchester | item management |
| ISO/IEC 14443 | Type A | Reader to Tag | 13.56 MHz | ASK | Modified Miller | Proximity cards, |
| | | Tag to Reader | 13.56 MHz | OOK | Manchester | item management |
| | Type B | Reader to Tag | 13.56 MHz | ASK | NRZ | |
| | | Tag to Reader | 13.56 MHz | BPSK | NRZ-L | |
| ISO/IEC 18092 (NFC) | A | Reader to Tag | 13.56 MHz | ASK | Modified Miller | Near-field communications |
| | | Tag to Reader | 13.56 MHz | ASK, OOK | Manchester | |
| | B | Reader to Tag | 13.56 MHz | ASK | NRZ | |
| | | Tag to Reader | 13.56 MHz | ASK, BPSK | NRZ | |
| | V | Reader to Tag | 13.56 MHz | ASK | PPM | |
| | | Tag to Reader | 13.56 MHz | ASK,OOK,FSK | Manchester | |
| ISO 21007 (HF) | - | - | 13.56 MHz | ASK | Miller | Identification of gas cylinders |
| ISO/IEC 18000-7 | - | - | 433.92 MHz | FSK | Manchester | Container/pallet tracking and security |
| ISO 18185-5 | Type A | Long-range | 433 MHz | FSK | Manchester | Electronic seals of freight |
| | | Short-range | 123–125 kHz | OOK | Manchester | containers and other supply |
| | Type B | Long-range | 2.45 GHz | BPSK | Differential | chain applications |
| | | Short-range | 114–126 kHz | FSK | Manchester | |
| ISO/IEC 18000-6 | Type A | Reader to Tag | 860–960 MHz | ASK | PIE | Large item management, |
| | | Tag to Reader | 860–960 MHz | ASK | FM0 | vehicle identification, |
| | Type B | Reader to Tag | 860–960 MHz | ASK | Manchester | supply chain, access/security |
| | | Tag to Reader | 860–960 MHz | ASK | FM0 | |
| ISO 18000-6C (EPC Class 1 Gen 2) | - | Reader to Tag | 860–960 MHz | DSB/SSB/ PR-ASK | PIE | Item management, vehicle identification, |
| | | Tag to Reader | 860–960 MHz | ASK or FSK | FM0, Miller | supply chain, access/security |
| ISO 10374 | - | - | 860–960 MHz, 2.45 GHz | FSK | Manchester | Identification of freight containers |
| ISO/IEC 18000-4 | Mode 1 | Reader to Tag | 2.45 GHz | ASK | Manchester | Road tolls, large item |
| | | Tag to Reader | 2.45 GHz | ASK | FM0 | management, supply chain, |
| | Mode 2 | Reader to Tag | 2.45 GHz | GMSK | None | access/security |
| | | Tag to Reader | 2.45 GHz | DBPSK or OOK | Manchester | |

- Sniff and emulate. The last step of the methodology is a trial and error process that requires to sniff and emulate communications to perform security tests. Sniffing is not only useful for reverse engineering a communications protocol, but also when trying to understand a well-documented standard protocol. Eventually, once the communications protocol is understood, it may be emulated with the appropriate hardware. For instance, Proxmark 3 official firmware offers off-the-shelf emulation of different standards (i.e., ISO/IEC 14443-A and 14443-B, ISO/IEC 15693) and specific tags (e.g., iClass, MIFARE, HID, Hitag, EM410x, Texas Instruments LF tags, or T55XX transponders). In the case of other platforms, an implementation of the reverse-engineered protocol may be necessary. For example, two cases of UHF RFID tag emulation using an USRP platform are presented in [85,86].

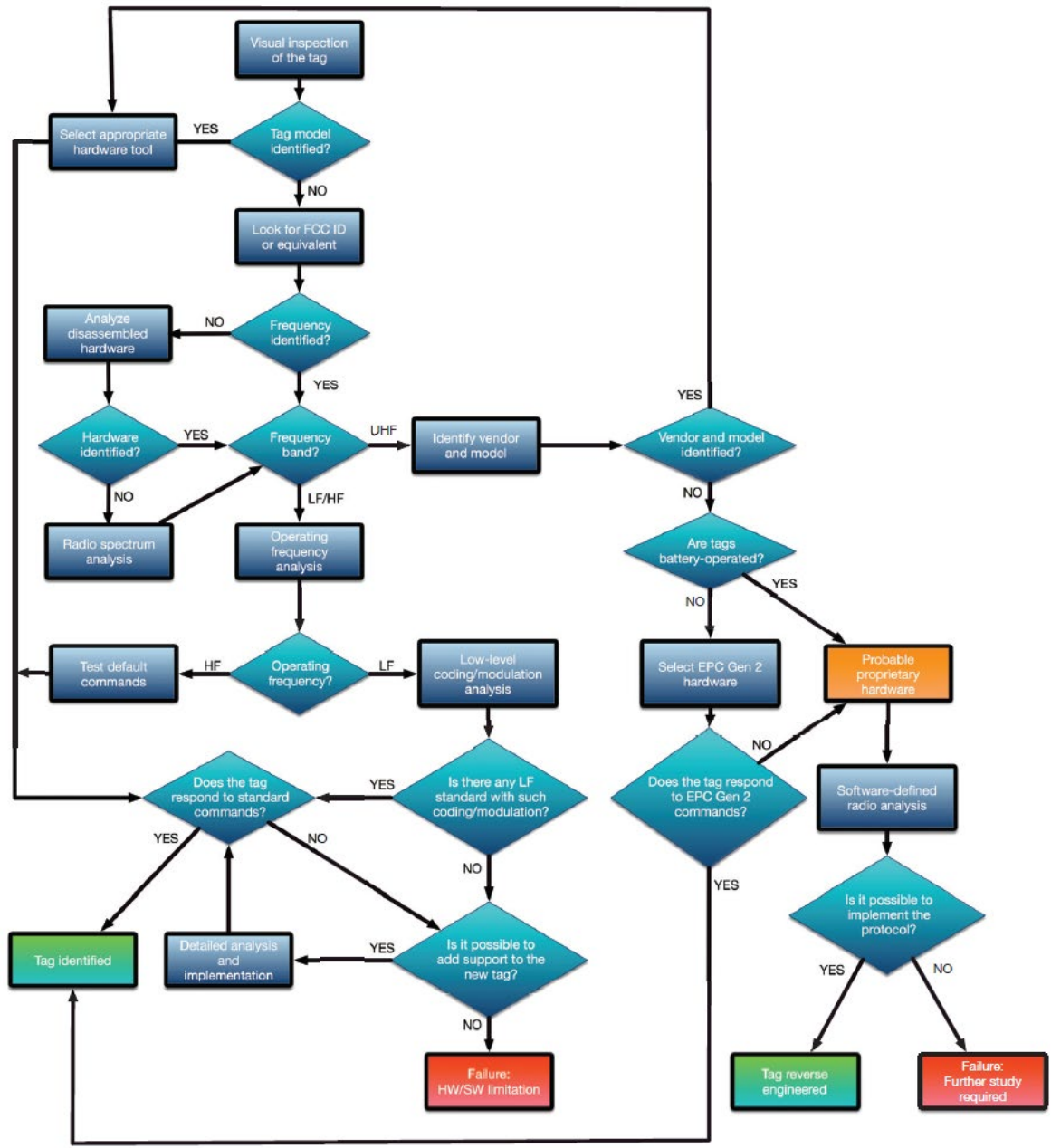

**Figure 3.** Flow diagram of the methodology.

### 3.3. A Practical Approach to the Methodology Using Proxmark 3

Although security has been improved in RFID systems over the last years, many commercial HF and LF tags currently deployed have not taken care of it properly. Thus, the methodology proposed can be easily applied with the help of tools like Proxmark 3 to a great deal of practical applications in the field of access control, transportation systems or supply chain tracking. However, it is important to note that Proxmark 3, although able to cover most commercial RFID tags, which work in LF or HF, it does not support operating in the UHF band.

Note that RFID UHF systems have their own peculiarities when securing them. First, they usually have longer reading ranges than LF and HF systems, so their communications can be intercepted and jammed from further distances. In passive systems, this is true mainly for the reader-to-tag channel, since communications tag-to-reader are restricted by the low-power electronics of the tags (therefore, tag-to-reader communications are harder to sniff). Note also that the increased power associated with UHF long range communications usually requires more powerful and expensive readers than in LF and HF RFID.

Perhaps the main limitation when assessing UHF RFID systems is the fact that they are not as standardized as in the LF and HF bands. The only relevant initiatives for a global standard have been carried out for the EPCglobal and the ISO 18000-6 tags. Nonetheless, while these specifications were not established, many companies pushed their own implementations, what derived in the fact that, in the UHF band, reverse engineering and security vary a lot depending on the manufacturer. In the same way, since the reader and tag hardware also varies substantially among manufacturers, in practice, researchers have to use generic SDR systems like the ones cited in the previous subsection (i.e., USRP, MyriadRF, HackRF One), but taking into account that the communities behind them are not as focused on RFID like in the case of Proxmark 3.

Regarding LF and HF tag analyses with Proxmark 3, they have to be approached in a different way, since their low-level behavior varies noticeably: as it will be described in the next subsections, it is possible to study LF tags easily at a physical-layer level, but it is not so easy in the case of HF devices. Due to this fact, the methodology distinguishes between both types of tags and requires different steps for each one.

### 3.3.1. Detecting the Operating Frequency

After an unsuccessful visual inspection of the tag, the methodology indicates that the operation frequency has to be obtained. When using the Proxmark 3, the operation frequency can be determined by first placing one of the antennas (LF or HF) far from the tag analyzed and then executing the command hw tune. A sequence diagram that illustrates the inner workings of the Proxmark 3 when executing such a command is presented in Figure 4. The sequence begins with the execution of the command, which sends a request to the Proxmark 3 ARM microcontroller through the USB. Next, the microcontroller asks the FPGA for the ADC values when tuning the LF and HF antennas to different frequencies. Eventually, the voltages associated with such frequencies are obtained and presented to the user. In the case of LF, the optimal resonant frequency is also estimated.

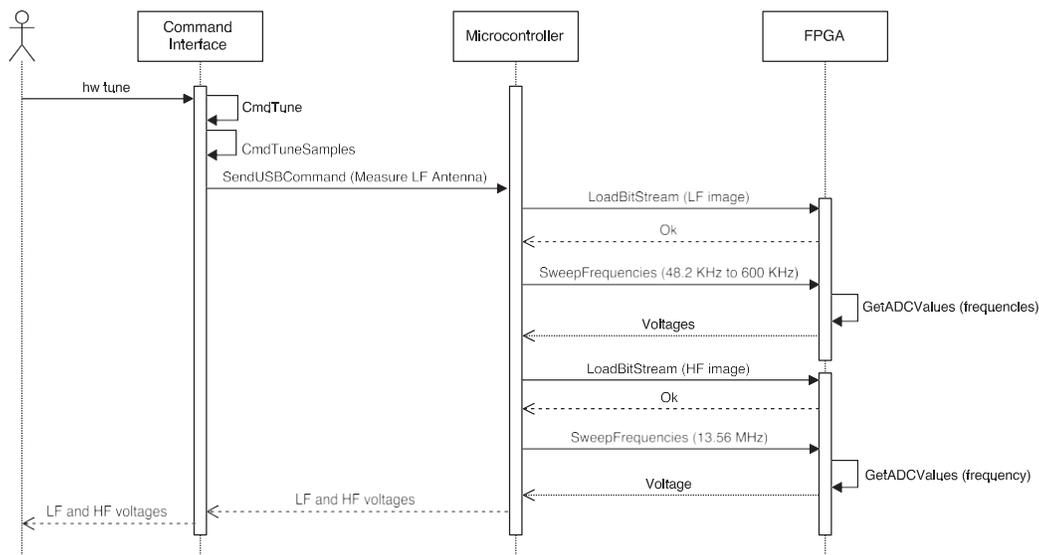

**Figure 4.** Sequence diagram of the command hw tune.

In order to determine the influence of the RFID tag analyzed on the reader (i.e., the Proxmark 3), the same operation has to be repeated next to such a tag: the operation frequency will be the one where the Proxmark 3 indicates that the voltage has dropped remarkably. If one of the antennas (HF or LF) does not show any changes in voltage for all the frequencies, it must be replaced by the other one and the same steps previously described have to be carried out again.

Figures 5 and 6 show the operation frequency detection process when testing an LF tag. First, in Figure 5), the voltage for every frequency is measured with the HF antenna connected (note that in such a figure the LF antenna is said to be unusable), initially with the tag out of the reading range (first execution of hw tune) and next, with the LF tag close to the HF antenna. As it can be observed by comparing the output for the two commands, the voltage barely changes (i.e., just around 0.4 V, it decreases from 9.24 V to 8.87 V). When the same procedure is carried out with the LF antenna, the voltages associated with LF frequencies drop substantially (see Figure 6 at 134 kHz, where voltage falls from 22.61 V to 14.34 V). This allows us to conclude that the tag is indeed LF.

**Figure 5.** HF voltages for an LF tag when is present (second command) or not in the field.

```
proxmark3> hw tune
#db# Measuring complete, sending report back to host

# LF antenna: 0.00 V @   125.00 kHz
# LF antenna: 0.00 V @   134.00 kHz
# LF optimal: 0.00 V @ 12000.00 kHz
# HF antenna: 9.24 V @    13.56 MHz
# Your LF antenna is unusable.

proxmark3> hw tune
#db# Measuring antenna characteristics, please wait..
#db# Measuring complete, sending report back to host
# LF antenna: 0.00 V @   125.00 kHz
# LF antenna: 0.00 V @   134.00 kHz
# LF optimal: 0.00 V @ 12000.00 kHz
# HF antenna: 8.87 V @    13.56 MHz
# Your LF antenna is unusable.
```

**Figure 6.** LF voltages when an LF tag is not in the field (first command) and when it is.

```
proxmark3> hw tune
#db# Measuring antenna characteristics, please wait..
#db# Measuring complete, sending report back to host

# LF antenna: 11.34 V @   125.00 kHz
# LF antenna: 22.61 V @   134.00 kHz
# LF optimal: 25.40 V @   131.87 kHz
# HF antenna:  0.62 V @    13.56 MHz
# Your HF antenna is unusable.

proxmark3> hw tune
#db# Measuring antenna characteristics, please wait..
#db# Measuring complete, sending report back to host

# LF antenna: 10.20 V @   125.00 kHz
# LF antenna: 14.34 V @   134.00 kHz
# LF optimal: 17.68 V @   139.53 kHz
# HF antenna:  0.67 V @    13.56 MHz
# Your HF antenna is unusable.
```

### 3.3.2. LF Tag Analysis

After determining that a tag works in the LF band, the methodology suggests figuring out its modulation and coding scheme. These tasks can be performed by following the next sequence of Proxmark 3 commands:

1.  LF read [h]: the tag is powered at the selected frequency (125 kHz by default, or 134 kHz using the optional parameter *h*). The command also records the signal transmitted by the tag.
2.  Data sample x: it downloads *x* of the previously recorded samples to the PC.
3.  Data plot. It allows the user to open a new window to plot the signal. It is useful for evaluating the signal visually.
4.  Different instructions can be used to modify, amplify, decimate or normalize signal values to ease signal identification.

5.  If the signal is clean enough and its modulation has been recognized, the user can try to demodulate it. For instance, if the signal is modulated in ASK, the command askdemod can be executed. In the case of FSK modulated signals, fskdemod is the right command.

An example of the output signal obtained after executing the previous three steps is illustrated in Figure 7. The demodulation command is shown in Figure 8, where 'X' characters are used to hide the actual UID. The command askdemod is first executed in such a figure, since at first sight the signal seems to be modulated in ASK, but an error was returned indicating that an ASK-modulated signal had not been detected. After taking a closer look at Figure 7, it could be observed that the period of the pulses with less amplitude is different from the others. Therefore, when fskdemod was executed, the signal was demodulated successfully. Figure 9 illustrates the sequence of steps performed by Proxmark 3 to demodulate an FSK signal successfully.

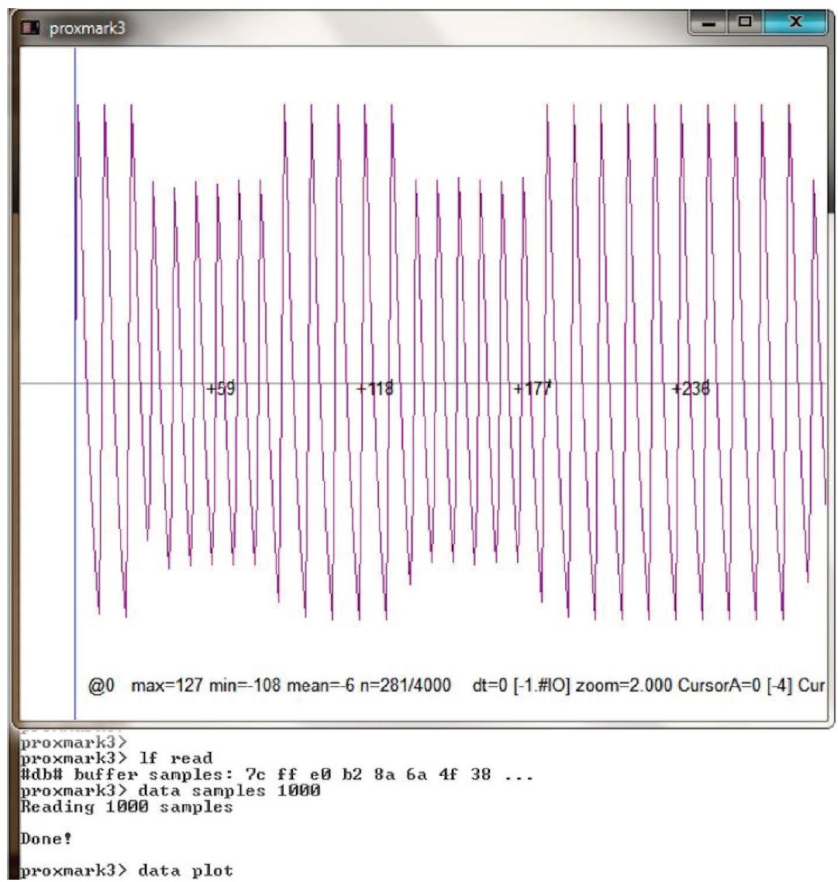

**Figure 7.** Example of an LF tag signal received.

```
proxmark3> data fskdemod
actual data bits start at sample 3212
length 50/50
bits: '010010101000001011000100100000000XXXX1001XXXX'
hex: 00000950 58900X9X
```

**Figure 8.** LF tag signal demodulated with fskdemod.

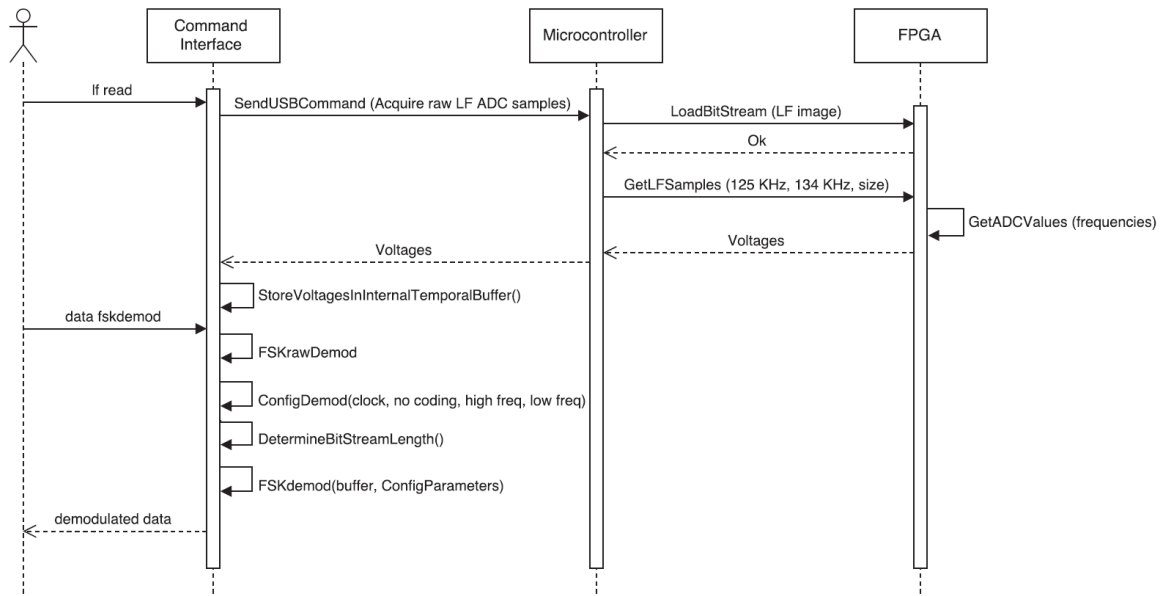

**Figure 9.** Sequence diagram of the commands executed for demodulating successfully an FSK signal.

After demodulation, decoding must be performed. It generally consists in looking for a bit pattern, which might lead to determine the length of the identifier transmitted. Thus, the signal has to be observed during certain periods of time and look for similarities. In order to understand the data transmitted, it can be useful to find the standard that defines and structures them. For instance, in the example illustrated in the previous figures, the LF tag was an access control card manufactured by HID [87], whose well-known LF data structures can be extracted and the UID can be obtained (as it is shown in Figure 10, where 'X' characters are used again to hide the real UID). The inner workings of Proxmark 3 during the execution of this command are almost identical to the ones shown in Figure 9, but signals are only acquired at 125 kHz and Manchester decoding is applied.

```
proxmark3> lf hid fskdemod
#db# TAG ID: 95058900X9X (1096)
```

**Figure 10.** Obtaining the tag UID of an access control LF tag manufactured by HID.

This specific HID tag can be emulated with the Proxmark 3 by using the command lf hid sim, and it can even be cloned with a re-writable tag like Atmel T5557.

### 3.3.3. HF Tag Analysis

The study of HF tags is different from the LF ones, since their signal is so fast that it cannot be analyzed easily at plain sight. Additionally, HF tags are generally smarter than LF tags, what allows them to perform more complex communications with the reader. There also exist many HF transmission modes and protocols, being HF tags and readers able to use several of them during the same transmission (for instance, a tag can send FSK-modulated data while the reader responds in ASK).

The steps required to analyze HF tags are not as clear as in the LF band, so the study becomes more like a trial-and-error process. An example is illustrated in Figure 11, where the data of an RFID card was decoded after trying one by one all the possible combinations defined by the most popular standards: first it was tested ISO/IEC 15693, then ISO/IEC 14443-A and, finally, ISO/IEC 14443-B. In this last case the command for reading tags [88] sends an ATQB command (0x05, 0x00, 0x08, 0x39, 0x73) and records the tag's answer. According to the standard, the second value of the output can be either 0x00000000 or 0x00000001 (if it is "1", it means the reply from the tag was received properly). If it is "0", it means that not all bytes (or none) were received.

```
proxmark3> hf 15 reader
#db# 0 octects read from IDENTIFY request:
#db# 0 octects read from SELECT request:
#db# 0 octects read from XXX request:

proxmark3> hf 14a reader
iso14443a card select failed

proxmark3> hf 14b read
#db# 3 1 e
```

**Figure 11.** Determining the RFID standard of an HF tag.

In the specific case of the previous tag, the answer was "3 1 e", so the second value ("1") means that the tag is actually compliant with ISO/IEC 14443-B. Figure 12 shows a simplified sequence diagram of the successful detection of an ISO/IEC 14443-B tag through Proxmark 3.

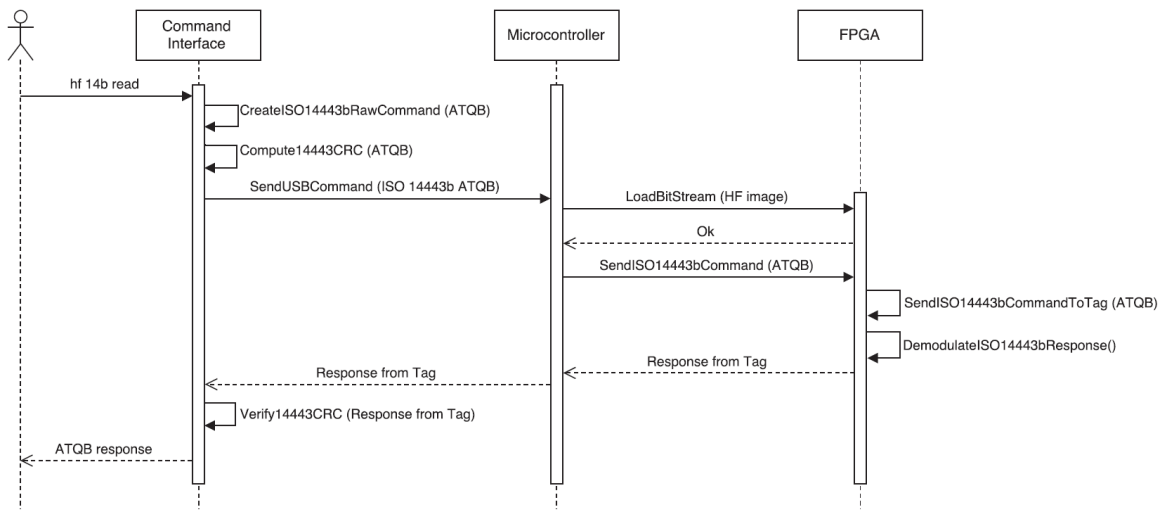

**Figure 12.** Simplified sequence diagram of the successful identification of an ISO/IEC 14443-B tag.

Furthermore, it would be possible for the Proxmark 3 to return the data after issuing the command hexsamples, which shows the UID and additional control bytes (in Figure 13).

```
proxmark3> data hexsamples
50 08 5c XX 7e XX 4f 44
4b 33 22 XX 74 XX 44 44
```

**Figure 13.** UID and control bytes from an ISO/IEC 14443-B compliant card.

## 4. Practical Evaluation

In order to validate the methodology proposed, three different commercial RFID tags were analyzed and tested. Such tags illustrate three different scenarios where secure access control and identification are required. Note that, although the systems analyzed were designed for very specific scenarios, the underlying RFID technology can be used in multiple IoT applications. The next subsections first introduce the tags audited and then give details on the analysis and the steps required to test their security.

### 4.1. HID Proximity Cards

HID Proximity are RFID cards used for access control. They are commonly used for accessing restricted areas without requiring keys.

### 4.1.1. Visual Inspection and FCC ID

In plain sight there are no signs or symbols that indicate the frequency band of the RFID card (see Figure 14). No FCC ID or other similar identifiers are included on the tag.

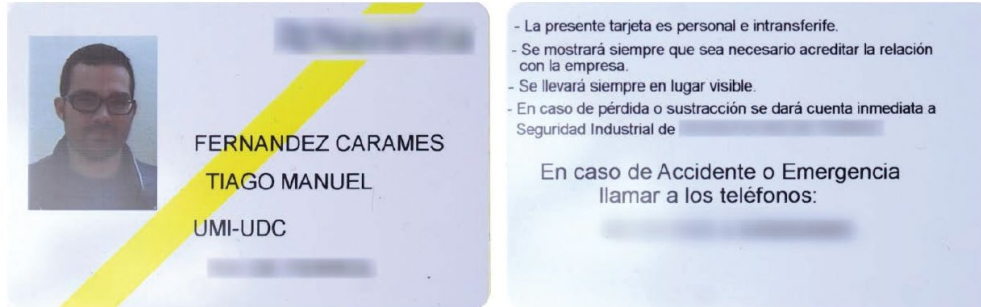

**Figure 14.** HID proximity card.

### 4.1.2. Analysis of Disassembled Hardware and Radio Spectrum Analysis

In this specific case, since there was just an RFID card available and no access to the reader, it was not possible to disassemble the hardware. What we can do is to take a shortcut and avoid using a spectrum analyzer: since the actual reading range is up to a few cm, the system is almost certainly HF or LF, what can be directly tested by using Proxmark 3 commands.

### 4.1.3. Operating Frequency and Modulation

- Radio frequency. The steps described in Section 3.3.1 were carried out and determined that it was an LF tag.
- Modulation. Once the radio frequency was obtained, a visual analysis of the signal received was performed to determine the modulation of the tag. Figure 15 allows us to conclude that the signal was modulated in FSK. In fact, it used a center frequency ($fc$) of 125 kHz and two sub-frequencies, *fc/8* and *fc/10*. Note that the signal shown in Figure 15 may seem initially an ASK wave, but after zooming in, it can be observed that the time period of the wider waves is larger than the one of the shorter waves.

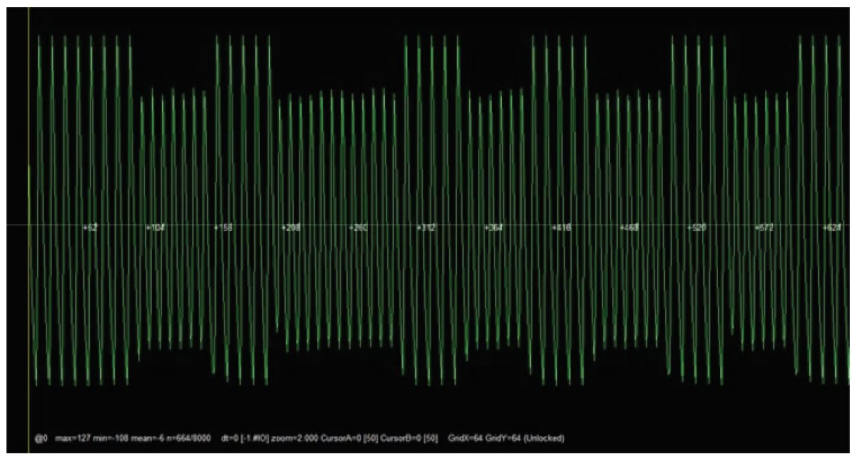

**Figure 15.** FSK modulated signal of the HID Proximity card.

### 4.1.4. Understanding the Underlying Protocols

HID Proximity card data are coded according to the following structure:

- The header follows a fixed pattern: 12 high-frequency (*fc/8*) pulses are first followed by a "0", and then 10 low-frequency (*fc/10*) pulses are followed by another "0".

- A "0" is coded with 5 low-frequency pulses followed by 5 high-frequency pulses.
- A "1" is coded with 6 high-frequency pulses followed by 6 low-frequency pulses.
- Every 4 data bits a high-frequency pulse is inserted.
- There are 44 data bits structured as indicated in Figure 16.

All these data can be read easily with the Proxmark 3 command fskdemod or by using a specific function offered by the official firmware that reads continuously the card IDs detected (lf hid fskdemod, mentioned previously in Section 3.3.2).

| Manufacturer Code | Card format / length (Card data) | Facility/Site Code | Card ID |
|---|---|---|---|
| 7 bits | 11 bits + 1 control bit | 8 bits | 16 bits + 1 control bit |

**Figure 16.** Data structure of an HID Proximity card.

### 4.1.5. Security Analysis

Security measures are almost non-existent in this kind of tags, which send continuously their ID. This lack of security is a problem, because their main use is for accessing buildings, certain rooms or restricted areas.

HID offers more secure systems, but, the system analyzed in this section is one of the most popular ones because is less expensive. The Proxmark 3 requires less than a second to read the ID and then it can reproduce it with the commands "lf sim" or "lf hid sim".

It is also possible to use a special firmware for Proxmark 3 developed by McAfee called ProxBrute [70]. Such a firmware allows attackers to apply brute force to the ID generation process. This is possible because all tags manufactured for a specific facility share the same Site Code, so the search space is reduced considerably: instead of generating combinations of 44 bits (17.5 trillions), only 26 bits have to be generated (67 million combinations).

Once the ID is known, the tag can be cloned. This can be achieved by using a T55x7 programmable tag and through the use of the Proxmark command "lf hid clone".

### 4.2. University of A Coruña's RFID Card

Until recently, the University of A Coruña gave students and staff an RFID card (in Figure 17) that was used for accessing different buildings and paying for different services (e.g., at the campus restaurants).

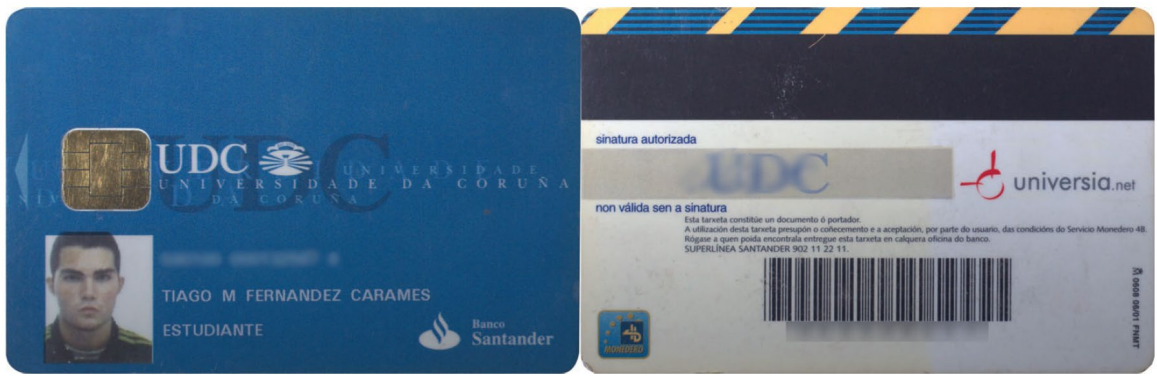

**Figure 17.** RFID card used in the University of A Coruña.

### 4.2.1. Visual Inspection and FCC ID

There is no external sign that identifies the underlying RFID technology. We can only see the contacts of traditional smart card interfaces. Thus, we can conclude that there are at least two interfaces:

one wired and another wireless. No FCC ID or equivalent can be observed on the tag, and we had no access to the official readers.

### 4.2.2. Analysis of Disassembled Hardware and Radio Spectrum Analysis

Like in the case of the HID card, we had just an RFID card available and no access to the reader, so it was not possible to disassemble the hardware. Similarly, we avoided performing a detailed analysis with the spectrum analyzer, since, due to the reading range (up to a few cm), it is fair to assume that the system was HF or LF, what can be directly tested by using the Proxmark 3 commands.

### 4.2.3. Operating Frequency and Modulation

- Operating frequency. The verification steps described in Section 3.3.1 allowed us to conclude that it was an HF card.
- Modulation. Once determined the frequency band, it was possible to test the commands for the different ISO/IEC standards. After testing the ones for ISO/IEC 14443-B and ISO/IEC 15693, it was found that the tag responded correctly to ISO/IEC 14443-A commands, which indicated that the tag was a MIFARE Classic 1K.

### 4.2.4. Understanding the Underlying Protocols

MIFARE is a contactless smartcard technology from NXP Semiconductors [89], that has sold more than 5 billion tags and fifty million RFID readers. It started to be manufactured around 1994–1995, being its first major deployment performed in Seoul's city transportation.

MIFARE is compliant with the first three parts of ISO/IEC 14443-A at 13.56 MHz, although there are certain differences depending on the tag version, which has been evolving over the last years.

MIFARE Classic is probably the most popular version of MIFARE cards. These tags use a really simple Application-Specific Integrated Circuit (ASIC) that basically stores data. Their memory is divided into sectors and blocks that are protected with a simple access control system. Each sector is divided into four blocks: three of them contain data, while the other one stores the data access permissions and the access keys.

There is not a fixed data format, although there is a special format called *value block* with specific operations for increasing and decreasing values. Sectors use two keys (A and B). Each key allows for managing different permissions: a key could be valid only for reading data, while the other one could be dedicated to modify them. The first 16 bytes of the internal memory are read-only and contain the serial number and other data related to the model and the manufacturer. Data are coded in Crypto-1, an already broken cryptographic protocol [42–44].

There are different MIFARE Classic versions:

- MIFARE Classic 1K. Its name derives from its 1024-byte internal storage, which is divided into 16 64-byte sectors.
- MIFARE Classic 4K. It has 4096 bytes for data divided into 40 sectors.
- MIFARE Classic Mini. It stores 320 bytes in 5 sectors (the actual useful data space is 224 bytes).

After MIFARE Classic, NXP created other versions: Ultralight, Ultralight C, DESFire (whose security was broken in 2011 [45]), Plus, DESFire V1/V2...

### 4.2.5. Security Analysis

As explained in the previous subsection, MIFARE Classic cards implement a security system that prevents reading or writing the internal data. However, this system is outdated and has already been broken.

To get the access keys to read and write the different internal blocks, the Proxmark official firmware offers several options. For instance, the command "hf mf mifare" executes the darkside attack [44] to obtain a valid key. The key to the attack is the authentication process, which is illustrated in Figure 18, where $n_T$ is a random 4-byte answer from the tag, $n_R$ is the 4-byte nonce chosen by the reader, suc is a bijective function, and $(ks_1, ks_2, ks_3)$ is a 96-bit keystream produced by the Crypto-1 stream cipher after being initialized with $n_T$ and $n_R$. The Dark Side attack exploits an implementation bug of Philips/NXP: it was found that, when the authentication process is run continuously using unknown keys, sometimes (1 out of 256 times), the card responded with 4 bits instead of 4 bytes when returning the value of $suc^3(n_T) \oplus ks_3$. These 4 bits are the Negative Acknowledgment (NACK) used to encrypt the next 4 bits of the keystream $ks_3$. From such an observation, the author of the attack created an algorithm to retrieve $ks_3$ by using low complexity and fast brute force attacks. In fact, such an attack usually takes from 30 s to half an hour (in practice, when using the Proxmark 3 implementation, sometimes, it has to be executed several times).

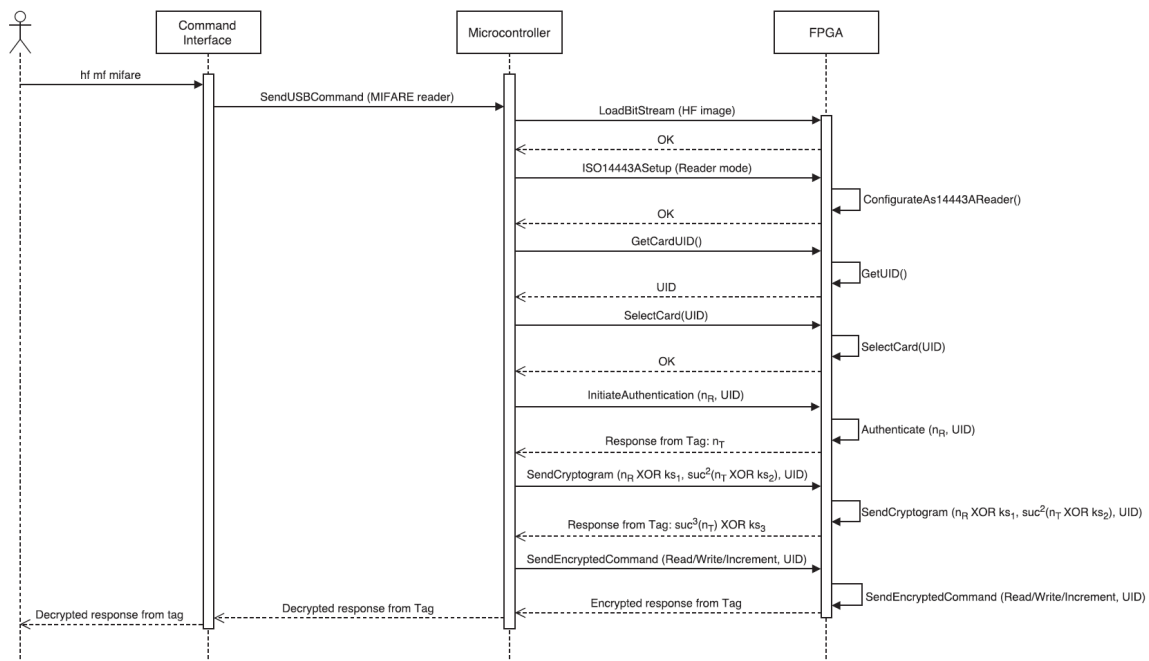

**Figure 18.** Authentication process of a MIFARE card in Proxmark 3.

An example of the system output after performing the attack is shown in Figure 19, where, after two unsuccessful attempts, the A key of the first block is finally obtained. Then, another attack called "nested authentication" [46] can be performed: it allows remote attackers to obtain the keys of all the other blocks (in Figure 20). Once all the keys have been obtained, a dump of the memory can be extracted.

With the dump it is possible to study the different parameters (e.g., detect memory changes as more payments are carried out) or save it to restore it later and recover the previous credit balance.

Figure 20 shows that, in the university card analyzed, the access keys are the ones used by default in every MIFARE Classic 1K. However, it is more critical its use for accessing restricted areas: the access is only controlled at the campus by the ISO/IEC 14443 UID, with no further verification of the card content, so it is extremely easy to clone the tags.

```
#db# COMMAND mifare FINISHED
key not found (lfsr_common_prefix list is null). Nt −c5823aa6
--------------------------------------------------------------------------------
Executing command. It may take up to 30 min.
Press the key on the proxmark3 device to abort both proxmark3 and client.

.............................#db# COMMAND mifare FINISHED
isOK:01

uid(0718b208) nt(26c732b8) par(179f5f0f67f7076f) ks(0f090c090101060e)

|diff|(nr)      |ks3|ks3^5|parity      |
+----+---------+---+-----+------------+
| 00 |00000000| f |  a  |1,1,1,0,1,0,0,0|
| 20 |00000020| 9 |  c  |1,1,1,1,1,0,0,1|
| 40 |00000040| c |  9  |1,1,1,1,1,0,1,0|
| 60 |00000060| 9 |  c  |1,1,1,1,0,0,0,0|
| 80 |00000080| 1 |  4  |1,1,1,0,0,1,1,0|
| a0 |000000a0| 1 |  4  |1,1,1,0,1,1,1,1|
| c0 |000000c0| 6 |  3  |1,1,1,0,0,0,0,0|
| e0 |000000e0| e |  b  |1,1,1,1,0,1,1,0|
key not found (lfsr_common_prefix list is null). Nt −26c732b8
--------------------------------------------------------------------------------
Executing command. It may take up to 30 min.
Press the key on the proxmark3 device to abort both proxmark3 and client.

..............................................
..............................................
...

isOK:01

uid(0718b208) nt(545371a7) par(3991710951990981) ks(0f01000b0a030a0f)

|diff|(nr)      |ks3|ks3^5|parity      |
+----+---------+---+-----+------------+
| 00 |00000000| f |  a  |1,0,0,1,1,1,0,0|
| 20 |00000020| 1 |  4  |1,0,0,0,1,0,0,1|
| 40 |00000040| 0 |  5  |1,0,0,0,1,1,1,0|
| 60 |00000060| b |  e  |1,0,0,1,0,0,0,0|
| 80 |00000080| a |  f  |1,0,0,0,1,0,1,0|
| a0 |000000a0| 3 |  6  |1,0,0,1,1,0,0,1|
| c0 |000000c0| a |  f  |1,0,0,1,0,0,0,0|
| e0 |000000e0| f |  a  |1,0,0,0,0,0,0,1|
#db# COMMAND mifare FINISHED
00ff9c86|00ffb8ed
--------------------------------------------------------------------------------
key found:ffffffffffff

Found valid key:ffffffffffff
```

**Figure 19.** Obtaining the first access key in a MIFARE Classic university card.

```
proxmark3> hf nf nested 1 0 a FFFFFFFFFFFF
-- block no:00 key type:00 key:ff ff ff ff ff
ff etrans:0
Block shift=0
Testing known keys. Sector count=16
nested...
Iterations count: 0
|---|------------------|---|------------------|---|
|sec|key A             |res|key B             |res|
|---|------------------|---|------------------|---|
|000| ffffffffffff     | 1 | ffffffffffff     | 1 |
|001| ffffffffffff     | 1 | ffffffffffff     | 1 |
|002| ffffffffffff     | 1 | ffffffffffff     | 1 |
|003| ffffffffffff     | 1 | ffffffffffff     | 1 |
|004| ffffffffffff     | 1 | ffffffffffff     | 1 |
|005| ffffffffffff     | 1 | ffffffffffff     | 1 |
|006| ffffffffffff     | 1 | ffffffffffff     | 1 |
|007| ffffffffffff     | 1 | ffffffffffff     | 1 |
|008| ffffffffffff     | 1 | ffffffffffff     | 1 |
|009| ffffffffffff     | 1 | ffffffffffff     | 1 |
|010| ffffffffffff     | 1 | ffffffffffff     | 1 |
|011| ffffffffffff     | 1 | ffffffffffff     | 1 |
|012| ffffffffffff     | 1 | ffffffffffff     | 1 |
|013| ffffffffffff     | 1 | ffffffffffff     | 1 |
|014| ffffffffffff     | 1 | ffffffffffff     | 1 |
|015| ffffffffffff     | 1 | ffffffffffff     | 1 |
|---|------------------|---|------------------|---|
```

**Figure 20.** Nested attack for an university RFID card.

*4.3. European Animal Identification Tags*

Since the late 1990s animal identification has been carried out throughout Europe. There are different kinds of tags, but one of the most common models is the one shown in Figure 21, which we have studied previously [53] when implanted subcutaneously. By means of similar tags, the members of the European Union track animal health of the most common pets, including cats, dogs and ferrets (European Regulation 998/2003). The same system is used in Europe for breeding and production of equidae (European Regulation 504/2008), and for public health in ovine and caprine animals (European Regulation 21/2004).

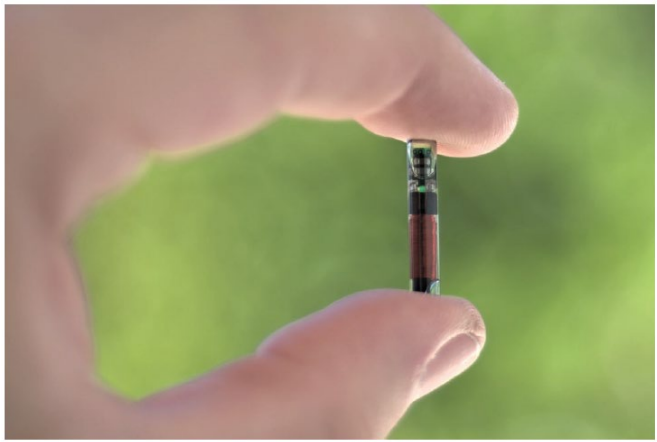

**Figure 21.** One of the animal identification tags analyzed.

4.3.1. Visual Inspection and FCC ID

These steps of the methodology are unnecessary, since these kinds of tags are regulated and specified by the different European regulations previously mentioned. We can jump directly to the study of the standards.

4.3.2. Detailed Analysis

European Regulation 998/2003 specifies that pet tags have to be compliant with ISO/IEC 11784 [90] and ISO/IEC 11785 [91]. They both describe LF tags, existing two different versions: half-duplex (HDX) and full-duplex (FDX and FDX-B). In Spain, FDX-B tags with bi-phase encoding are the most common.

- Operating frequency. With the help of the Proxmark 3 a tag implanted in a dog was verified. As expected, it was determined that it was LF.
- Modulation. It was not straightforward to recognize the modulation, since the signals received were very noisy (it must be noted that the tag had been implanted a year before the tests were performed). Figure 22 illustrates the noise level on the signals received. Due to such noise, it was necessary to filter the signal, obtaining a figure like the one shown in Figure 23, which resembles the expected bi-phase encoding.

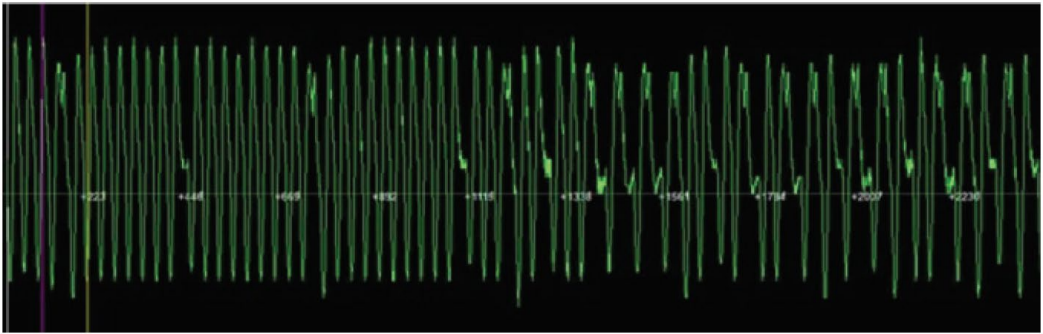

**Figure 22.** Noisy signal from an animal identification tag.

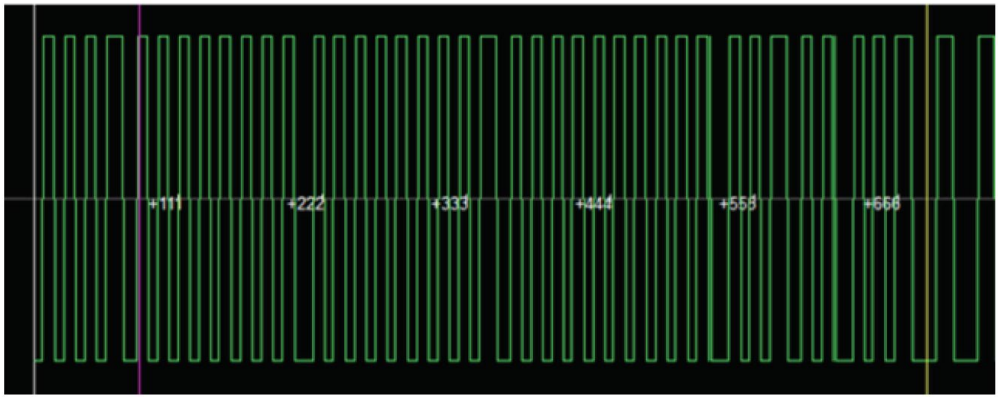

**Figure 23.** Animal identification tag signal after filtering it.

When these experiments were carried out, the official Proxmark firmware did not support FDX-B, so it was necessary to implement it. Such an implementation first filters and demodulates the signal, and then decodes it.

### 4.3.3. Understanding the Underlying Protocols

ISO/IEC 11784 and ISO/IEC 11785 are international standards that regulate RFID for animal identification. Each animal transponder contains 64 bits with the information shown in Figure 24 (the bit data values included were generated randomly).

| Field\#bit | 11 | 10 | 9 | 8 | 7 | 6 | 5 | 4 | 3 | 2 | 1 |
|---|---|---|---|---|---|---|---|---|---|---|---|
| | (msb) | | | | | | | | | | (lsb) |
| Header | 1 | 0 | 0 | 0 | 0 | 0 | 0 | 0 | 0 | 0 | 0 |
| National Code (38 bits) | | | 1 | 1 | 0 | 1 | 0 | 0 | 0 | 0 | 0 |
| | | | 1 | 0 | 1 | 0 | 0 | 1 | 1 | 1 | 1 |
| | | | 1 | 0 | 1 | 0 | 0 | 1 | 0 | 0 | 1 |
| | | | 1 | 0 | 1 | 0 | 1 | 0 | 1 | 0 | 1 |
| Country Code (10 bits) | | | 1 | 1 | 1 | 0 | 0 | 0 | 0 | 0 | 0 |
| | | | 1 | 1 | 1 | 0 | 0 | 0 | 1 | 1 | 1 |
| Data Block Status Flag (1 bit) | | | 1 | - | - | - | - | - | - | - | 1 |
| Animal Application Indicator (1 bit) | | | 1 | 1 | - | - | - | - | - | - | - |
| Checksum (16 bits) | | | 1 | 1 | 1 | 0 | 1 | 1 | 1 | 1 | 0 |
| | | | 1 | 0 | 0 | 1 | 0 | 1 | 0 | 1 | 1 |
| Optional Extra Data (24 bits) | | | 1 | 0 | 1 | 1 | 1 | 0 | 1 | 1 | 0 |
| | | | 1 | 1 | 1 | 1 | 0 | 0 | 0 | 0 | 1 |
| | | | 1 | 0 | 1 | 0 | 0 | 1 | 0 | 0 | 0 |

**Figure 24.** Internal memory structure of an animal identification tag.

The standard defines two different transmission modes at 134.2 kHz: Half-Duplex (HDX) and Full-Duplex (FDX or FDX-B). Since in HDX mode the tag is not able to send data and receive power at the same time, the process of reading requires to power up the tag for a short interval and then wait for the tag to transmit data. In HDX mode the header consists of 8 bits (always "01111110") and the Cyclic-Redundancy Check (CRC) is 16-bit long. In addition, a 24-bit chunk of data is sent with information on the application. All data are modulated in FSK and coded with NRZ.

In FDX-B mode tags are able to transmit data and be powered at the same time. Figure 24 shows the internal data structure defined by the standard: it includes an 11-bit header ("10000000000"), 50 bits of data, 24-bits with the application information and a 16-bit CRC. Moreover, a control bit is added (always "1") every 8 bits (except for the header). Data are sent in Less-Significant Bit (LSB) order, so, when the reader receives the bits, it can reconstruct them just by using simple binary shifts. These bits are modulated in ASK and coded in DBP.

### 4.3.4. Security Evaluation

Before performing the security evaluation, it was necessary to implement the appropriate FDX-B functions. Such an implementation required the following tasks (illustrated in Figure 25 as a sequence diagram):

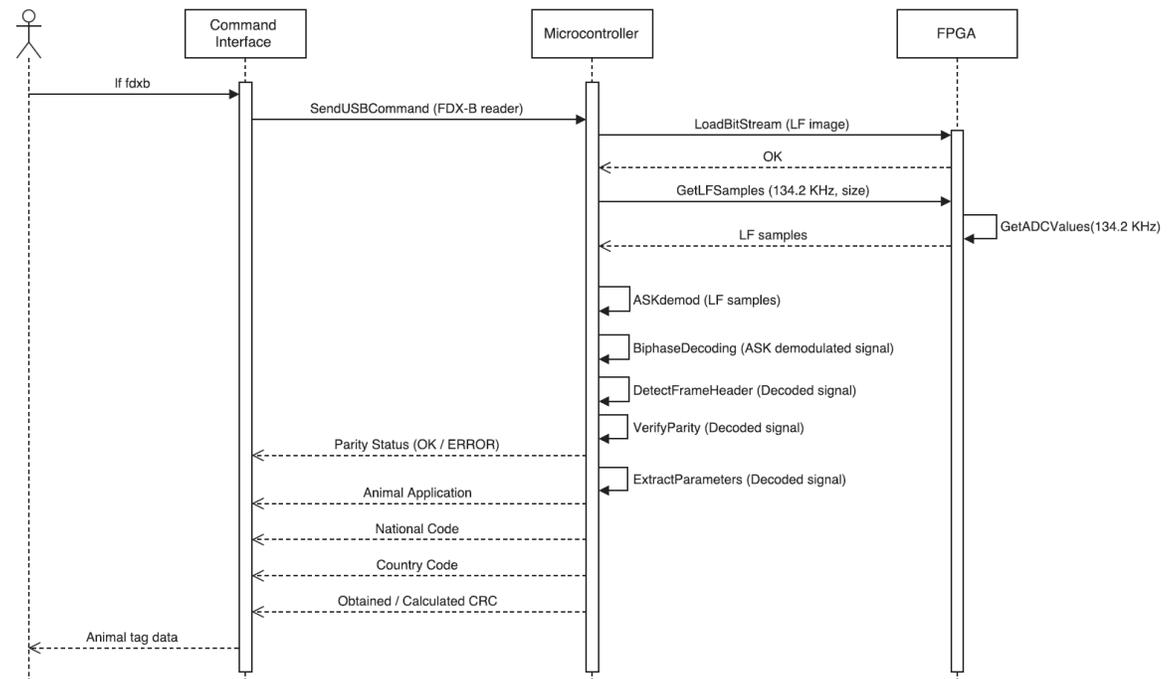

**Figure 25.** Sequence diagram of the Proxmark 3 FDX-B implementation.

- Modify the original Proxmark 3 client application (programmed in C) to offer users a new function to perform FDX-B demodulation.
- Modify the original Proxmark 3 microcontroller firmware (also programmed in C) to create an FDX-B demodulation function that first demodulates the ASK signal and then decodes the DBP bit stream. It was not necessary to reprogram the FPGA firmware. However, this FDX-B demodulation function requires to manage the input/output and the exchange memory buffers used by the already-available FPGA functions to perform the operations as fast as possible.
- ASK demodulation was already available in the original firmware, but it was necessary to modify the Proxmark 3 microcontroller firmware to implement the DBP decoding function.

After implementing the FDX-B functions, it was straightforward to read data from any FDX-B tag. The software developed extracts the two main parameters: the country code and the national code (the actual identifier). Figure 26 shows an example where two consecutive readings were performed: the first one shows errors related to a bad reading (almost certainly caused by the noise), while the second one was successful.

```
Header found, starting data in pos 279
Bit control error CC in bit 332
Bit control error in bit 341
Bit control error in bit 359
Bit control error in bit 368
Animal APP
National code: 193452859733
Country code: 249
Obtained CRC: aae1
Calculated CRC: 21ab

Header found, starting data in pos 147
Animal APP
National code: 098104278921
Country code: 981
Obtained CRC: 3763
Calculated CRC: 3763
```

**Figure 26.** Example of readings from an animal identification tag.

It can be concluded that security is almost non-existent in these kinds of tags. Although writing on the tags is not allowed, they continuously send the stored data without any authentication requirement. It may seem that the application is not susceptible for including high-security mechanisms, since its main objective is to identify the health records and the owner of an animal, but, in terms of privacy and uniqueness of the identifier, the current system is not effective. Note that, using a device like Proxmark 3, it is not only easy to read the data, but also to emulate tags and clone them.

What might worry users is that these tags can be attached to animals aimed at producing human food (i.e., sheep and goats). Cloning or erasing the tag data breaks traceability, which is the way to determine where an epidemic outbreak is originated. European Union authorities should take these risks into account and implement security measures to preserve traceability. Some authors have proposed solutions to improve traceability trustworthiness, being some of the latest ones related to the application of blockchain technology [5].

## 5. Conclusions

RFID is one of the key technologies for the development of IoT applications, but it is important to take security into consideration to avoid privacy and security risks. This article included three main contributions aimed at fostering security in RFID-based IoT applications. First, it presented a detailed review of some of the latest and most common attacks on RFID systems. Such a review was completed with a clear description of the most recent hardware and software RFID security tools. Second, due to the lack of a step-by-step methodology for auditing RFID communications security, a novel approach was presented. Third, the application of such a methodology was illustrated through three real-world applications where relevant flaws were detected. The tests performed have shown that, by using a device like Proxmark 3 and a minimum of reverse engineering skills, it is possible to clone animal identification information, to alter data of payment cards, to extract private information from certain cards, to capture tag-reader communications, and to emulate both readers and tags.

The final conclusion is that, although many applications can make use of advanced security RFID measures, certain developers have adopted the technology without taking such mechanisms into account. In the case of the access control tags analyzed, their security can be improved by adding a higher security layer (e.g., encrypting internal data), enabling some of the already existing security protocols, or simply replacing the tags with more secure versions. In the case of animal identification, more sophisticated measures should be implemented due to the criticality of the system in terms of human health.

To sum up, a methodology like the one proposed in this paper can help IoT application developers to perform audits and determine the security level of an RFID system before taking it from a test environment to a real-world situation. Finally, note that, although this paper is aimed at fostering security in RFID-based IoT applications, the methodology proposed can be applied to any RFID system.

**Acknowledgments:** This work has been funded by the Spanish Ministry of Economy and Competitiveness under grants TEC2013-47141-C4-1-R and TEC2016-75067-C4-1-R.

**Author Contributions:** Tiago M. Fernández-Caramés and Paula Fraga-Lamas conceived and designed the experiments; Tiago M. Fernández-Caramés and Manuel Suárez-Albela performed the experiments; Paula Fraga-Lamas, Luis Castedo and Tiago M. Fernández-Caramés analyzed the data; Manuel Suárez-Albela, Paula Fraga-Lamas, Luis Castedo and Tiago M. Fernández-Caramés wrote the paper.

**Conflicts of Interest:** The authors declare no conflict of interest. The founding sponsors had no role in the design of the study; in the collection, analyses, or interpretation of data; in the writing of the manuscript, and in the decision to publish the results.

## Abbreviations

The following abbreviations are used in this manuscript:

| | |
|---|---|
| ADC | Analog-to-Digital Converter |
| ASIC | Application-Specific Integrated Circuit |
| ASK | Amplitude-Shift Keying |
| BPSK | Binary-Phase Shift Keying |
| CRC | Cyclic-Redundancy Check |
| DBP | Differential Bi-Phase |
| DBPSK | Differential BPSK |
| DP | Dual Pattern |
| DSB-ASK | Double-Sideband ASK |
| DoS | Denial of Service |
| FPGA | Field-Programmable Gate Array |
| FSK | Frequency-Shift Keying |
| GMSK | Gaussian Minimum Shift Keying |
| HF | High Frequency |
| ITS | Intelligent Transportation Systems |
| LF | Low Frequency |
| LSB | Less-Significant Bit |
| MAC | Medium Access Control |
| MitM | Man-in-the-Middle |
| MFM | Modified Frequency Modulation |
| NACK | Negative Acknowledgment |
| NRZ | Non-Return-to-Zero |
| NRZ-L | Non-Return-to-Zero-L |
| OOK | On-Off Keying |
| PIE | Pulse-Interval Encoding |
| PJM | Phase-Jitter Modulation |
| PPM | Pulse-Position Modulation |
| PR-ASK | Phase-Reversal ASK |
| RFID | Radio Frequency Identification |

| SHF | Super-High Frequency |
| SSB-ASK | Single-Sideband ASK |
| UHF | Ultra-High Frequency |
| UID | Unique Identifier |